\documentclass[aps,prd,twocolumn,superscriptaddress,nofootinbib,10pt]{revtex4-1}

\usepackage[utf8]{inputenc}
\usepackage{graphicx,amsmath,amsfonts,amssymb,aas_macros,slashed}
\usepackage{bbold}
\usepackage{datetime}
\usepackage{numprint}
\usepackage{enumitem}
\usepackage{booktabs}
 
\usepackage{graphicx,color,amsmath}
\usepackage{hyperref}

\allowdisplaybreaks

\newcolumntype{C}{>{$}c<{$}}
\AtBeginDocument{
\heavyrulewidth=.08em
\lightrulewidth=.05em
\cmidrulewidth=.03em
\belowrulesep=.65ex
\belowbottomsep=0pt
\aboverulesep=.4ex
\abovetopsep=0pt
\cmidrulesep=\doublerulesep
\cmidrulekern=.5em
\defaultaddspace=.5em
}

\newcommand{\keV}{\ensuremath{\mathrm{keV}}}
\newcommand{\MeV}{\ensuremath{\mathrm{MeV}}}
\newcommand{\GeV}{\ensuremath{\mathrm{GeV}}}

\usepackage{pgfplots}
\usepackage{bm}
\usepackage{mathtools}

\makeatletter
\newcommand{\thickhline}{%
    \noalign {\ifnum 0=`}\fi \hrule height 1pt
    \futurelet \reserved@a \@xhline
}
\newcolumntype{"}{@{\hskip\tabcolsep\vrule width 1pt\hskip\tabcolsep}}
\makeatother

\usetikzlibrary{positioning,arrows,patterns}
\usetikzlibrary{decorations.markings}
\usetikzlibrary{calc}

\tikzset{
    photon/.style={decorate, decoration={snake}, draw=black},
    vector/.style={decorate, decoration={snake}, draw},
	provector/.style={decorate, decoration={snake,amplitude=2.5pt}, draw},
	antivector/.style={decorate, decoration={snake,amplitude=-2.5pt}, draw},
    fermion/.style={draw=black, postaction={decorate},
        decoration={markings,mark=at position .55 with {\arrow[draw=black]{>}}}},
    fermionbar/.style={draw=black, postaction={decorate},
        decoration={markings,mark=at position .55 with {\arrow[draw=black]{<}}}},
    fermionnoarrow/.style={draw=black},
    gluon/.style={decorate, draw=black,
        decoration={coil,amplitude=4pt, segment length=5pt}},
    scalar/.style={dashed,draw=black, postaction={decorate},
        decoration={markings,mark=at position .55 with {\arrow[draw=black]{>}}}},
    scalarbar/.style={dashed,draw=black, postaction={decorate},
        decoration={markings,mark=at position .55 with {\arrow[draw=black]{<}}}},
    scalartwo/.style={dotted,draw=black, postaction={decorate},
        decoration={markings,mark=at position .55 with {\arrow[draw=black]{>}}}},
    scalartwobar/.style={dotted,draw=black, postaction={decorate},
        decoration={markings,mark=at position .55 with {\arrow[draw=black]{<}}}},
    scalarnoarrow/.style={dashed,draw=black},
    electron/.style={draw=black, postaction={decorate},
        decoration={markings,mark=at position .55 with {\arrow[draw=black]{>}}}},
	bigvector/.style={decorate, decoration={snake,amplitude=4pt}, draw},
    vertex/.style={draw,shape=circle,fill=black,minimum size=1pt,inner sep=0pt},
    fermion2/.style={double, draw=black, postaction={decorate},
		decoration={markings,mark=at position .55 with {\arrow[draw=black]{>}}}},
    momentum/.style={draw=black,line width=0.15mm, postaction={decorate},
        decoration={markings,mark=at position 1 with {\arrow[draw=black]{>}}}}
}

\begin{document}

\title{Terrestrial Probes of Electromagnetically Interacting Dark Radiation}

\author{Jui-Lin Kuo}
\email{jui-lin.kuo@oeaw.ac.at}
\affiliation{Institute of High Energy Physics, Austrian Academy of Sciences, Nikolsdorfergasse 18, 1050 Vienna, Austria}
\author{Maxim Pospelov}
\email{pospelov@umn.edu}
\affiliation{School of Physics and Astronomy, University of Minnesota, Minneapolis, MN
55455, USA}
\affiliation{William I. Fine Theoretical Physics Institute, School of Physics and
Astronomy, University of Minnesota, Minneapolis, MN 55455, USA}
\author{Josef Pradler}
\email{josef.pradler@oeaw.ac.at}
\affiliation{Institute of High Energy Physics, Austrian Academy of Sciences, Nikolsdorfergasse 18, 1050 Vienna, Austria}

\begin{abstract}
We study the possibility that dark radiation, sourced through the decay of dark matter in the late Universe, carries electromagnetic interactions. The relativistic flux of particles induces recoil signals in direct detection and neutrino experiments through its interaction with millicharge, electric/magnetic dipole moments, or anapole moment/charge radius. Taking the DM lifetime as 35 times the age of the Universe, as currently cosmologically allowed, we show that direct detection (neutrino) experiments have complementary sensitivity down to $\epsilon\sim 10^{-11}$ $(10^{-12})$,  $d_\chi/\mu_\chi \sim 10^{-9}\,\mu_B$ $(10^{-13}\mu_B)$, and   $a_\chi/b_\chi \sim 10^{-2}\,\GeV^{-2}$ $(10^{-8}\,\GeV^{-2})$ on the respective couplings. Finally, we show that such dark radiation can lead to a satisfactory explanation of the recently observed XENON1T excess in the electron recoil signal without being in conflict with other bounds.
\end{abstract}

\maketitle

\section{Introduction}
\label{sec:introduction}

The identity of dark matter (DM) in the Universe remains to be a mystery. While the gravitational manifestations of DM are numerous and well studied, the connection between DM and the Standard Model (SM) of particles and fields is unknown. A large amount of effort and resources have been invested in the attempts to detect the non-gravitational interaction of SM particles with DM. Thus far, these efforts have resulted in stringent limits on the strength of interaction for certain types of DM. In particular,  DM in the form of individual particles with masses comparable to the masses of SM particles has been constrained~\cite{Bauer:2013ihz}. While many of these searches were targeting weakly interacting massive particles (or WIMPs), over time it has become clear that the sensitivity of the existing experiments extends beyond WIMP-nucleon scattering and beyond the electroweak scale for DM masses, see {\em e.g.} \cite{Pospelov:2008jk,Essig:2011nj,Graham:2012su,Kouvaris:2016afs,Hochberg:2016sqx,Ibe:2017yqa}.

In parallel to the experimental developments, the last two decades brought a more general view on DM physics. Early on ({\em e.g.}~with the example of supersymmetric WIMP relics \cite{Ellis:1983ew}) it was understood that the DM may not come ``in isolation'', but be, in fact, a member of a more generic  dark sector. 
This sector could comprise additional heavy particles charged under the SM gauge group ({\em e.g.} weak scale supersymmetry), or alternatively may include the mediators of interaction or carriers of a ``dark force'', as well as new very light degrees of freedom known as dark radiation (DR). Light particles from dark sectors would almost invariably have to have a small coupling to the SM. Dark forces and light mediators have received significant attention in the literature, both in the cosmological and laboratory settings~\cite{Alexander:2016aln,Battaglieri:2017aum,Tulin:2017ara,Beacham:2019nyx,Lanfranchi:2020crw}.
In comparison, dark radiation has mostly been studied in the context of its contribution to the cosmological expansion rate, parametrized by $N_{\rm eff}$. 
One interesting aspect of DR (for a set of representative ideas see Refs.~\cite{Hasenkamp:2012ii,Buen-Abad:2015ova,Ko:2016fcd,Cui:2017ytb,Pospelov:2018kdh,Bringmann:2018jpr,McKeen:2018pbb,McKeen:2018xyz,Chacko:2018uke,Bondarenko:2020vta,Berghaus:2020ekh,Dror:2021nyr,Jaeckel:2021gah})  is that it can be created in a non-thermal way, and therefore differ in energy from the characteristic energies of the quanta of the cosmic microwave background. In this paper we will discuss some observable signatures of interacting DR in the regime when $\omega_{\rm DR} \gg \omega_{\rm CMB}$ holds for the respective typical energies.

The goal of the present paper is to consider a class of dark radiation models that interacts with the SM via a single photon exchange. 
We will consider DR originating from the late decays of DM particles with mass $m_{X} >  {\rm keV}$. This range will allow to probe such scenarios using DM direct detection experiments, and for $\omega_{\rm DR}$ above 200\,keV, using sensitive underground neutrino experiments. The first goal of this project is to map the sensitivity of the best existing experiments vs.~the strength of DR electromagnetic form factors controlling the interaction and available DR fluxes. The latter depend on the mass and lifetime of decaying progenitor particles. 

The second goal of this paper is to consider DR and the multitude of possible electromagnetic form factors as candidates for the explanation of the recently reported signal excess in the XENON1T experiment \cite{Aprile:2020tmw}. The excess, consistent with the injection of $O(2-3)$\,keV electromagnetic energy, has numerous candidate explanations. The collaboration itself has tried to connect it with DR coming from the Sun, in form of the axions and/or neutrinos with electromagnetic dipole interaction. Our goal is then to generalize it to a number of possible form factors, in the situation when the DR radiation flux is maximized by employing the DM$\to$DR decay. It is unclear if the explanation of this excess can be achieved with DR without being in conflict with other measurements and constraints.

We will adopt a fairly minimal scheme, where the progenitor decay, $X\to \bar \chi \chi$, is sourcing DR in form of $\chi$'s, which we assume to be a Dirac fermion. 
The Lagrangians specifying the~$\chi$  interactions with the photon $A_{\mu}$ or its field strength tensor $F_{\mu\nu}$ in ascending order of their dimensionality read,
\begin{subequations}
  \label{eq:Lagrangians}
\begin{align} 
 \mathcal{L}_{\rm dim = 4} & =   \epsilon e \, \bar\chi \gamma^{\mu} \chi A_{\mu}  , \\
\mathcal{L}_{\rm dim = 5} & =   \frac{1}{2} \mu_\chi \, \bar\chi \sigma^{\mu\nu} \chi F_{\mu\nu}    + \frac{i}{2} d_\chi \, \bar\chi  \sigma^{\mu\nu}\gamma^5 \chi F_{\mu\nu} , \\
\mathcal{L}_{\rm dim = 6} & =  - a_\chi \,  \bar\chi \gamma^{\mu} \gamma^5\chi \partial^{\nu} F_{\mu\nu}  + b_\chi \,  \bar\chi \gamma^{\mu} \chi \partial^{\nu} F_{\mu\nu} .
\end{align}
\end{subequations}
Here $\epsilon e $ is the millicharge (mQ), $\mu_\chi$ and $d_\chi$ are the magnetic and electric dipole moments (MDM and EDM), and $a_\chi$ and $b_\chi$ are the anapole moment and charge radius interaction (AM and CR), respectively. If $\chi$ were a single Majorana fermion, then only the $a_\chi$ interaction is allowed. If $\chi$ is instead a complex scalar, $\mu_\chi$, $d_\chi$ and $a_\chi$ will vanish. For a real scalar such form factors simply do not exist. Thus, we consider DR with the Dirac fermion case to have the most variety of phenomenological consequences. In the past, various aspects of (effective) electromagnetic couplings of dark sector particles (including constraints from direct detection, beam dump limits, SM precision observables, colliders and stellar constraints) were explored in a number of publications \cite{Pospelov:2000bq,Sigurdson:2004zp,Barger:2010gv,Ho:2012bg,Schmidt:2012yg,Kopp:2014tsa,Ibarra:2015fqa,Sandick:2016zut,Kavanagh:2018xeh,Trickle:2019ovy,Chu:2018qrm,Chu:2019rok,Chang:2019xva,Chu:2020ysb,Marocco:2020dqu,Arina:2020mxo}; for the mQ interaction, see the recent review~\cite{Lanfranchi:2020crw} and references therein.

The paper is organized as follows: In Sec.~\ref{Sec:flux}, we compute the DR flux from decaying DM (DDM), followed by an overview of experiments in Sec.~\ref{Sec:exp}.
In Sec.~\ref{Sec:event_rate}, we derive the expected electron recoil (ER) or nuclear recoil (NR) event rate by the DR. 
In Sec.~\ref{sec:result}, we demonstrate the constraints and the forecasts of sensitivity on the parameter space and present the fit to the XENON1T excess.
Finally, in Sec.~\ref{sec:conclusions}, we conclude and give outlooks.

\section{Dark radiation flux}
\label{Sec:flux}

In this section, we collect the ingredients for the calculation of the DR flux from DDM. 
For simplicity, we consider a two-body decay $X \rightarrow \bar \chi \chi $ and assume a single decay channel for~$X$. In that case, $X\bar\chi\chi$ coupling can be traded for the lifetime of $X$. 
The expected Galactic energy differential flux is given by 
\begin{equation}
\label{Eq:galactic_differential_flux}
\dfrac{d \phi_\chi^{\rm gal}}{dE_\chi} = \dfrac{e^{-t_0 /\tau_X}}{m_X \tau_X} \dfrac{dN_\chi}{dE_\chi} R_\odot \rho_\odot \langle D \rangle\,,
\end{equation}
where $t_0$ is the age of the universe, $m_X$ and $\tau_X$ are the DM mass and lifetime, $R_\odot \simeq 8.33\,{\rm kpc}$ is the distance between the Sun and the Galactic center,  $\rho_\odot = 3\times 10^5 \,{\rm keV}/{\rm cm}^3$ is the local DM energy density and $\langle D \rangle \simeq 2.1$ is the averaged $D$-factor assuming a NFW profile~\cite{Navarro:1995iw}. For simplicity, we assume a 100\% decaying fraction of DM; if this is not the case, the formulas are to be dressed with the DDM fraction in an obvious way.
The DR injection spectrum is monochromatic,
\begin{equation}
   \dfrac{dN_\chi}{dE_\chi} = 2 \delta\left( E_\chi - \dfrac{m_X}{2}	 \right)\,,
\end{equation}
with a negligible spread by the parent DM velocity dispersion. 

In turn, the energy differential DR flux that originates from DDM cosmologically reads~\cite{Cui:2017ytb}, 
\begin{equation}
	\dfrac{d \phi_\chi^{\rm ext}}{dE_\chi} = \dfrac{2\Omega_X \rho_c}{ m_X \tau_X H_0 p_\chi} \dfrac{e^{-t(\xi -1)/\tau_X}}{\sqrt{\xi^3 \Omega_m +\Omega_\Lambda}} \Theta(\xi -1)\,,
\end{equation}
where $\Omega_X = 0.2607$ is the DM density parameter~\cite{Aghanim:2018eyx} and $\rho_c = 4.82 \,{\rm keV}\,{\rm cm}^{-3}$ is the critical density of the Universe at present, $\Theta (\xi -1)$ is a Heaviside step function, and $\xi \equiv p_{\rm in}/ p_\chi$  is the ratio of injected momentum $p_{\rm in}$ to arriving momentum $p_\chi$, $p_{\rm in}^2 = (m_X/2)^2 - m_\chi^2$.
In the exponential $t(\xi - 1 )$ is the cosmic time at redshift $z = \xi -1$. For a spacially flat cosmology, for $z\lesssim 10^3$, it is given by
\begin{equation}
	t(z) = \dfrac{1}{3H_0 \sqrt{\Omega_\Lambda}} \ln \left[ \dfrac{\sqrt{1+(\Omega_m /\Omega_\Lambda) (1+z)^3}+1}{\sqrt{1+(\Omega_m /\Omega_\Lambda) (1+z)^3}-1}\right]\,,
\end{equation}
where $\Omega_m = 0.3111$, $\Omega_\Lambda = 0.6889$ are the cosmological density parameters for matter and dark energy, respectively; $H_0 = 67.66 \,{\rm km\,s^{-1}\, Mpc^{-1}}$ is our adopted present day Hubble rate~\cite{Aghanim:2018eyx}.

\begin{figure}[tb]
\centering
\includegraphics[width=\columnwidth]{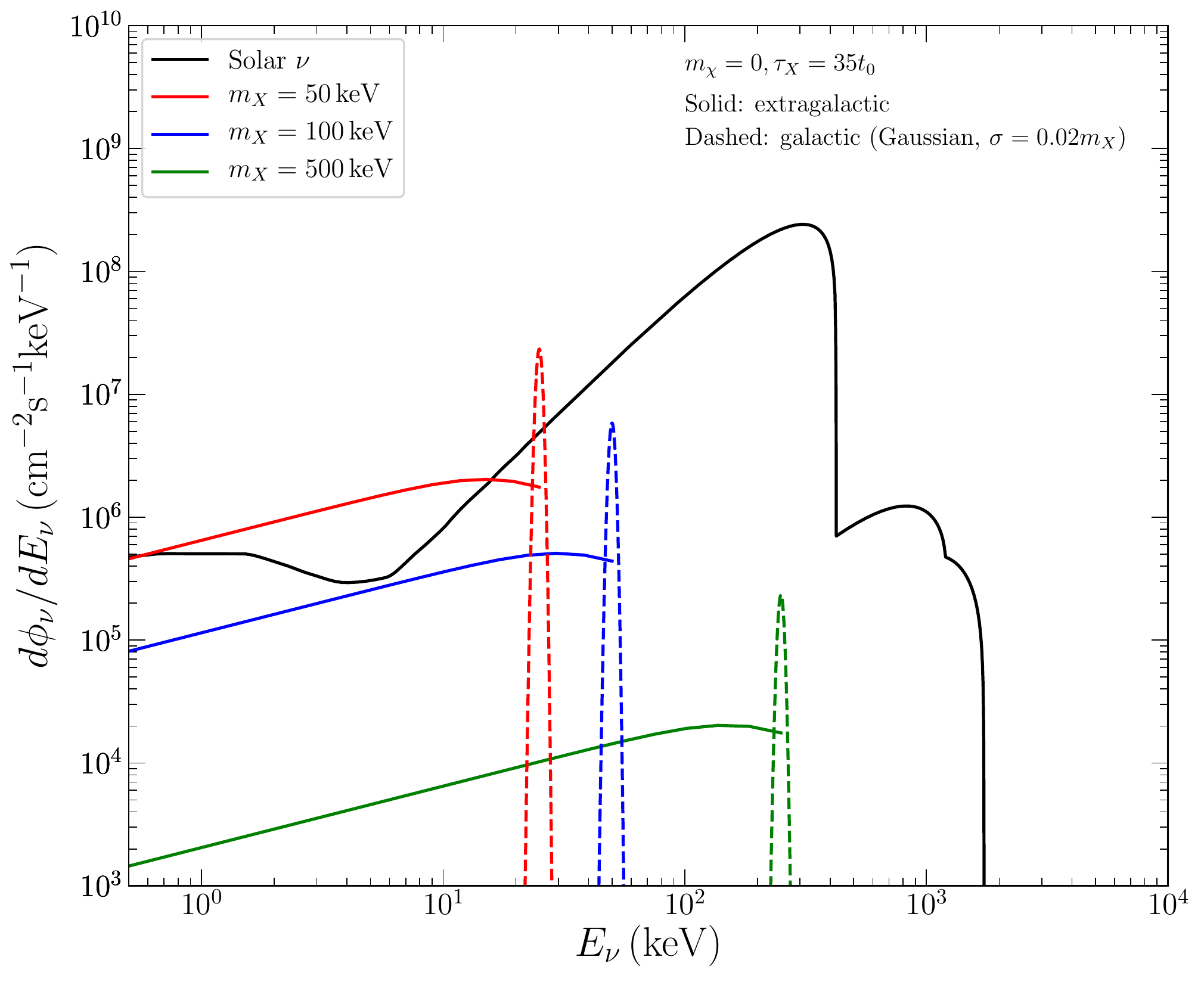}
\caption{The solar neutrino flux (solid black) and the DR flux from the DDM for  $\tau_X = 35t_0$ and various choices of $m_X$. Both, galactic and extragalactic DR flux from DDM can reach comparable levels of flux with respect to the solar neutrinos. 
}
\label{Fig:nu_flux}
\end{figure}

As benchmark value for the DM lifetime, we take $\tau_X = 35 t_0$~\cite{Chen:2020iwm}, which saturates the limit on invisibly DDM from a joint data set that includes cosmic microwave background measurements~\cite{Aghanim:2018eyx,Aghanim:2019ame}, the Pantheon data of type Ia supernovae~\cite{Scolnic:2017caz} and baryon acoustic oscillation measurements~\cite{2012MNRAS.423.3430B,Ross:2014qpa,Alam:2016hwk}; for previous constraints on $\tau_X$ or the fraction of DDM, see~\cite{Enqvist:2015ara,Poulin:2016nat,Nygaard:2020sow}. The DR mass is not entering the analysis in an appreciable way, as we focus on the relativistic daughter particles, unless the value of $m_\chi $ is explicitly stated.
Therefore, we have two free parameters: the DM mass $m_X$ and the coupling between the DR and the SM sector.

In Fig.~\ref{Fig:nu_flux}, we compare the solar neutrino flux and the expected galactic DR flux (dashed lines) and the extragalactic DR flux (solid lines) originating from DDM with $m_X = 50, 100 , 500\,{\rm keV}$; we apply a 2\% Gaussian smearing on the monochromatic Galactic flux for visualization.
The fluxes are compared to the solar neutrino flux (solid black line) taken from~\cite{Bahcall:1987jc,Bahcall:1996qv,Bahcall:1997eg}; below 10~keV, we include the contribution from plasmon decay, photoproduction, and  bremsstrahlung from~\cite{Vitagliano:2017odj}.
As can be seen, both galactic and extragalactic DR fluxes are, in magnitude, in roughly the same ballpark as the solar neutrino flux. 

\section{Experiments}
\label{Sec:exp}

\begin{table*}[t]
 \centering
 \begin{tabular}{lcccr}
	 \toprule
    &Exposure (ton$\times$yr) & Signal Range & Signal Type  & Reference \\
  \midrule
  XENON1T (fit)  & $0.65$ & $[1,10]\,{\rm keV}$ & ER  & \cite{Aprile:2020tmw}  \\
  XENON1T (S1+S2)  & $0.65$ & $[3,66]\,{\rm PE_{S1}}$ & ER  & \cite{Aprile:2020tmw}  \\
    XENON1T (S2)  & $0.06$ & $[150,526]\,{\rm PE}$ & ER  & \cite{Aprile:2019xxb}  \\
   Borexino & $2.1\times10^2$ & $[0.32,2.64]\,{\rm MeV}$ & ER, NR  & \cite{Agostini:2020lci,Agostini:2020mfq} \\
   Super-Kamiokande & $9.2 \times 10^4$ & $[16,88]\,{\rm MeV}$ & ER  & \cite{Bays:2011si}  \\
   & $1.6 \times 10^5$ & $[0.1,1.33]\,{\rm GeV}$ & ER  &  \cite{Kachulis:2017nci} \\
  Hyper-Kamiokande$^*$ & $2.3 \times 10^6$ & $[16,88]\,{\rm MeV}$ &  ER & \cite{Abe:2011ts,Kearns:2013lea} \\
  & $4.0\times 10^6$ & $[0.1,1.33]\,{\rm GeV}$ & ER  & \cite{Abe:2011ts,Kearns:2013lea}  \\
  DUNE$^*$ (10/40\,kton) & $7.2 \times 10^4 \,(2.9 \times 10^5)$ & $[0.03,1.33]\,{\rm GeV}$ &  ER & \cite{Necib:2016aez} \\
  \bottomrule
 \end{tabular}
 \caption{Summary of experiments with (effective) exposures, our considered signal ranges, signal type, and main reference for the reported data used in this work; the star indicates that a forecast on the sensitivity is derived.}
  \label{tab:experiment}
 \end{table*}

In this section, we outline the considered experiments and the way to derive constraints and forecasts of sensitivity on the parameter space. 
For ER in the $\mathcal{O}({\rm keV})$ energy ballpark, we consider the scattering of DR in the XENON1T detector.
Neutrino experiments such as Borexino, Super-Kamiokande (SK) as well as the future Hyper-Kamiokande (HK) and Deep Underground Neutrino Experiment (DUNE) have larger energy threshold, ${\rm MeV}\text{--}{\rm GeV}$ range,  and we consider  DR-electron scattering for which the solar neutrinos ($E_R < 30 \,{\rm MeV}$) and atmospheric neutrinos $(E_R > 30 \,{\rm MeV})$ become the main background.
For Borexino, we consider DR-proton scattering in addition.

\subsection{XENON1T}

The  XENON1T detector, located underground at the Gran Sasso laboratory, is a dual-phase time projection chamber with liquid and gaseous xenon. 
The registered signals include prompt scintillation (S1) and secondary scintillation from ionization (S2). 
In a recent analysis~\cite{Aprile:2020tmw}, an excess of events was identified in the S1 data at $\mathcal{O}({\rm keV})$. Although poorly understood backgrounds exist~\cite{Aprile:2020tmw,Szydagis:2020isq}, the possibility that this signal is due to new physics has been entertained abundantly, see~\cite{Aprile:2020tmw,Boehm:2020ltd,Takahashi:2020bpq,An:2020bxd,Khan:2020vaf,Bloch:2020uzh,DiLuzio:2020jjp,*Buch:2020mrg,*Gao:2020wer,*Dent:2020jhf,*Dessert:2020vxy,*Sun:2020iim,*Cacciapaglia:2020kbf,*Croon:2020ehi,*Li:2020naa,*Athron:2020maw,*Millea:2020xxp,*Arias-Aragon:2020qtn,*Alonso-Alvarez:2020cdv,*Choi:2020kch,*Chiang:2020hgb,*Kannike:2020agf,*Fornal:2020npv,*Su:2020zny,*Chen:2020gcl,*Cao:2020bwd,*Jho:2020sku,*DelleRose:2020pbh,*Alhazmi:2020fju,*Basu:2020gsy,*Davoudiasl:2020ypv,*Choudhury:2020xui,*Ema:2020fit,*Cao:2020oxq,*Bally:2020yid,*AristizabalSierra:2020edu,*Ge:2020jfn,*Coloma:2020voz,*Miranda:2020kwy,*Babu:2020ivd,*Shoemaker:2020kji,*Arcadi:2020zni,*Karmakar:2020rbi,*Harigaya:2020ckz,*Bell:2020bes,*Lee:2020wmh,*Bramante:2020zos,*An:2020tcg,*Chao:2020yro,*Baek:2020owl,*He:2020wjs,*Borah:2020jzi,*Du:2020ybt,*Choi:2020udy,*Paz:2020pbc,*Dey:2020sai,*Budnik:2020nwz,*Zu:2020idx,*Lindner:2020kko,*McKeen:2020vpf,*Gao:2020wfr,*Ko:2020gdg,*Okada:2020evk}.
The excess is not in conflict with an earlier S2-only analysis by the experiment~\cite{Aprile:2019xxb}.
Here we derive both, the favored region for the anomaly and the constraints on the parameter space using the S1+S2 and S2-only data.
Details on the limit-setting procedure can be found in~\cite{An:2020bxd} which we follow here; see also~\cite{An:2017ojc}.

\subsection{Borexino}

\begin{figure}[tb]
\begin{center}
\includegraphics[width=0.49\textwidth]{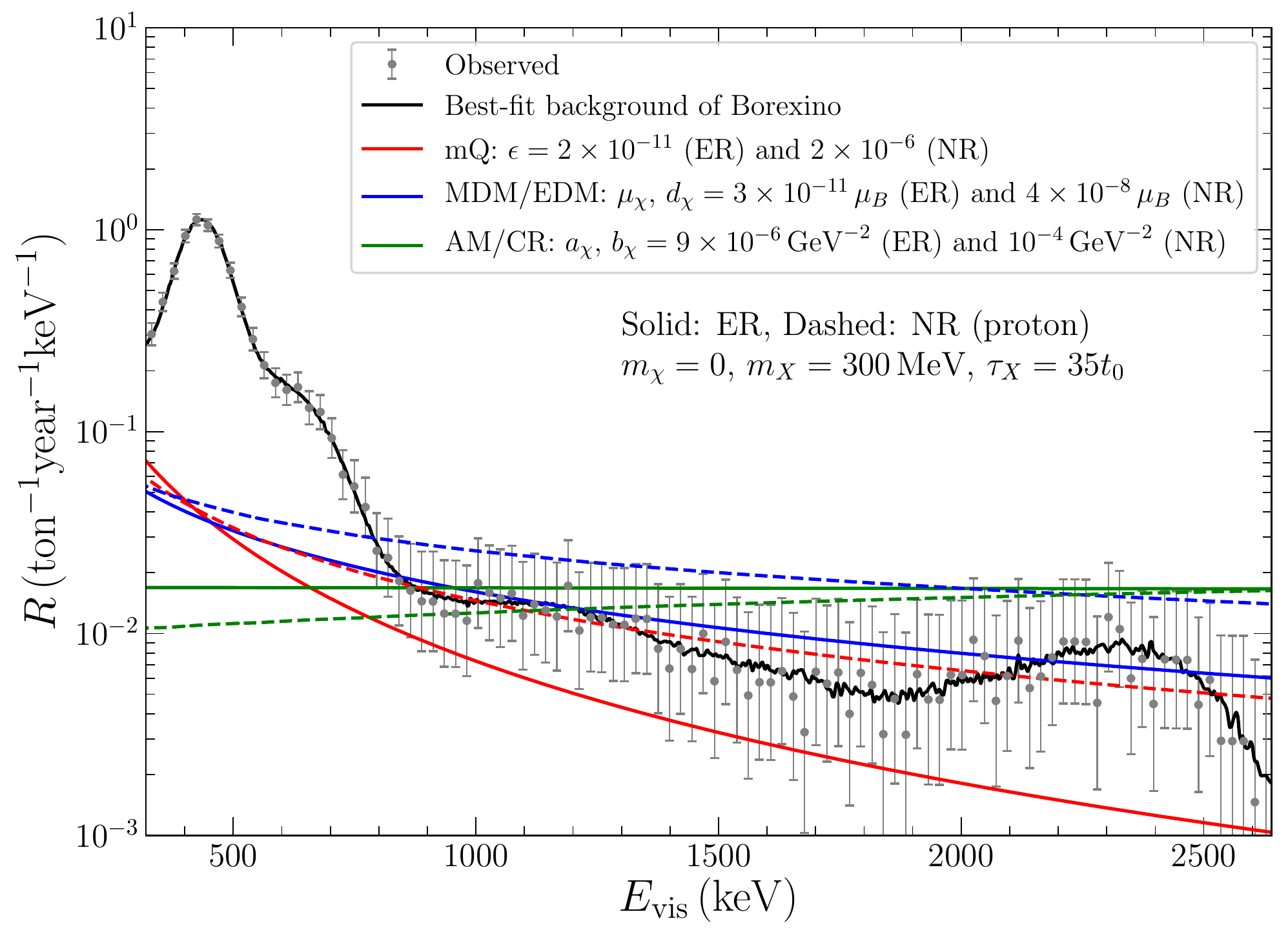}
\end{center}
\caption{The Borexino observed event rate together with the reported best-fit background model (solid black), and exemplary event rates for mQ (red), MDM/EDM (blue) AM/CR (green) as a function of the visible energy $E_{\rm vis}$. 
Solid curves are for ER, while dashed ones are for NR.
The differing energy-dependence of the operators of differing dimensionality can be clearly observed.  
}
\label{Fig:Borexino_spectrum}
\end{figure}

The Borexino experiment features a liquid scintillator-based detector with $280\,{\rm ton}$  fiducial mass, primarily designed to measure solar neutrinos in the quasi-elastic scattering signal with electrons~\cite{Alimonti:2008gc}.
We use the latest data from the CNO neutrino search of phase-III~\cite{Agostini:2020lci,Agostini:2020mfq} of the experiment, with an exposure of 209.4~${\rm ton}\text{-}{\rm yr}$.
Between the threshold energy $320\,{\rm keV}$ and $2640\,{\rm keV}$, the observed event rate and the best-fit background plus solar neutrino-induced rate are reported for each energy bin.
Note that the standard neutrino events  are a background in our consideration.
The detection efficiency is assumed to be unity.
We derive $95\%$ C.L. limits using the ${\rm CL}_s$ method~\cite{Read:2002hq}.

For heavier progenitor masses, we further consider the proton recoil signal in the Borexino detector. 
Here, we adopt Birk's law to account for the energy quenching in the organic scintillator,
\begin{equation}
   E_{\rm vis} = \int_0^{E_R} \dfrac{dE}{1+k_B dE/dx}\,,
\end{equation}
where $E_{\rm vis}$ is the visible energy, $k_B \simeq 0.01\,{\rm cm}/{\rm MeV}$ is Birk's constant and $dE/dx$ is the stopping power which we compute using the \texttt{SRIM} computer package; see also~\cite{Pospelov:2012gm}.
For the scintillator pseudocumine ${\rm C}_9 {\rm H}_{12}$ with a mass density of $\rho = 0.88\,{\rm g}/{\rm cm}^3$, the stopping power for protons is roughly $dE/dx \sim \mathcal{O}(100)\,{\rm MeV}/{\rm cm}$, albeit energy-dependent.
For electrons, $dE/dx \sim \mathcal{O}(10^{-3})\,{\rm MeV}/{\rm cm}$ so that we are allowed to neglect the energy quenching since $dE/dx \ll k_B^{-1}$. 
We utilize the same data and method presented above to derive the corresponding constraint. 
See Fig.~\ref{Fig:Borexino_spectrum} for a demonstration of the event rate from different operators and the Borexino data.

\subsection{Super-Kamiokande}
\label{Sec:SK}
Super-Kamiokande (SK) is a neutrino experiment with a water-based Cherenkov detector located $2.7\,{\rm km}$ underground in Japan.
The  fiducial mass is $22.5\,{\rm kton}$.
First, we consider the low-energy $e^-$-recoil data with $E_R = 16\text{--}88\,{\rm MeV}$ from a diffuse supernova neutrino background search in the SK-I run~\cite{Bays:2011si}.
With 1497 days of observation, 239 events are reported with $N_{\rm bkg} = 238$ from the best-fit model, which has also been utilized to set bounds on neutrino DR~\cite{Cui:2017ytb} and cosmic-ray upscattered DM~\cite{Cappiello:2019qsw}.
The corresponding efficiency is taken from~\cite{Bays:2011si}.
At the higher recoil energy range, a recent analysis~\cite{Kachulis:2017nci} provides three energy bins of $161.9\,{\rm kton}\text{-}{\rm yr}$ fiducialized fully-contained data from the SK-IV run, with  cuts applied for a single relativistic electron and no accompanying nuclear interaction.
We use the first energy interval ranging from $100\,{\rm MeV}$ to $1.33\,{\rm GeV}$ with a total number of  $N_{\rm obs} = 4042$ events and an efficiency $\epsilon (0.5\,{\rm GeV}) = 0.93$.
The estimated background is 3993 $e^-$-recoil events during its data-taking time~\cite{Kachulis:2017nci}.
The ensuing $90\%$~C.L.~limits on the various signal strengths can be derived by requiring that the DR-induced events $N_{\rm sig}^{\rm DR}$ satisfy, 
\begin{equation}
\label{Eq:90CL}
    N_{\rm sig}^{\rm DR} \leq {\rm Max}[0,N_{\rm obs} - N_{\rm bkg}] + 1.28 \sqrt{N_{\rm obs}}\, .
\end{equation}

\subsection{Hyper-Kamiokande}
Hyper-Kamiokande (HK) will be equipped with 25 times larger fiducial mass than SK~\cite{Abe:2011ts,Kearns:2013lea}. It will  provide supreme sensitivity to  solar, atmospheric and supernova neutrinos.
We consider the same low and high recoil energy ranges as in SK.
The background estimation is done by rescaling the background events of SK according to their difference in the fiducial mass.
Under the assumption of same data-taking time as SK and a constant efficiency of $0.8$, we derive the projected sensitivity of HK by imposing %
$N_{\rm sig}^{\rm DR} \leq 1.28 \sqrt{N_{\rm bkg}}$ assuming $N_{\rm obs} = N_{\rm bkg}$.

\subsection{DUNE}

DUNE is a proposed long-baseline neutrino facility, which serves as the far detector for the neutrino beam generated from $1300\,{\rm km}$ away~\cite{Abi:2017aow}.
As an add-on, its liquid argon (LAr)-based detector can also probe light dark sector physics~\cite{Acciarri:2015uup}.
DUNE will be comprised of four $10\,{\rm kton}$ detectors.
In the following we consider both the $10\,{\rm kton}$ and $40\,{\rm kton}$ configurations.
To avoid the immense solar neutrino background, the  electron energy threshold is set to $30\,{\rm MeV}$~\cite{Necib:2016aez}.
The expected (all-sky) number of $e^-$-recoil background events per year 
is  $N_{\rm bkg} = 128\, (512)$ for the $10\,(40)\,{\rm kton}$ detector~\cite{Necib:2016aez}.
The detection efficiency for the LAr time projection chamber is assumed to be 0.5.
Finally, we obtain the future projection on the  couplings for each progenitor mass by the condition $N_{\rm sig}^{\rm DR}\leq 1.28 \sqrt{N_{\rm bkg}}$, assuming $N_{\rm obs} = N_{\rm bkg}$ and the same data-taking time as well as the upper boundary of recoil energy as the SK high-$E_R$ data in Sec.~\ref{Sec:SK}.

\section{event rate}
\label{Sec:event_rate}

\subsection{Scattering on bound electrons}
For small progenitor mass, the resulting DR is low-energetic enough that we need to account for bound state effects in the DR-electron scattering and resulting atomic ionization process. 
Combining the DR flux from Sec.~\ref{Sec:flux} and the differential cross section given in App.~\ref{App:xsec_electron}, the differential event rate for scattering with the electrons is 
\begin{equation}
\label{Eq:differential_rate}
\dfrac{dR}{dE_R} = \kappa N_T  \varepsilon(E_R) \int_{q_-}^{q_+} dq \int_{p_\chi^{\rm min}}^{p_\chi^{\rm max}} dp_\chi \, \dfrac{p_\chi}{ E_\chi}\dfrac{d\phi_\chi}{dE_\chi} \dfrac{d \sigma v}{dq dE_R}\,,	
\end{equation}
where $\kappa$ is the exposure of the experiment, $N_T$ is the number of targets per detector mass, $ \varepsilon(E_R)$ is the detection efficiency, and $d\phi_\chi/ dE_\chi$ is the differential $\chi$ flux from DDM that includes both, the galactic and extragalactic components.
The minimum $\chi$-momentum for a given recoil energy $E_R$ and momentum transfer $q$ is 
\begin{equation}
\label{Eq:pchiminbound}
p_\chi^{\rm min} = \dfrac{q}{2 x} \left[ x + \dfrac{\Delta E}{q}\sqrt{x\left(x+ \dfrac{4m_\chi^2}{q^2} \right)}\right]\,,
\end{equation}
where $x = 1 - \Delta E^2/ q^2$ with $\Delta E = E_R + |E_B^{n,l}|$ being the deposited energy and $E_B^{n,l}$ is the binding energy of the bound state orbital $(n,l)$.
The upper boundary of the $p_\chi$ integration is given by $p_\chi^{\rm max} = p_{\rm in} $.
The integration boundaries of $q$ are given by
\begin{equation}
q_+ = p_\chi^{\rm max} + \dfrac{\sqrt{(m_X - 2 \Delta E)^2 -4 m_\chi^2}}{2}\,,\,\, q_- = \Delta E\,.
\end{equation}
 To obtain the total event rate, we sum up the contributions from all kinematically available $(n,l)$ shells.

\subsection{Scattering on free particles}
For larger progenitor mass, the $\mathcal{O}({\rm MeV}\text{--}{\rm GeV})$ ER signals induced by DR are best probed in the large-volume neutrino experiments mentioned in Sec.~\ref{Sec:exp}.
For such recoil energies, the initial electron can be considered as a free particle.
With the recoil cross section given in Appendix.~\ref{App:xsec_free}, the total differential event rate reads
\begin{equation}
    \dfrac{dR}{dE_R} = \kappa  N_T \varepsilon(E_R) \int_{p_\chi^{\rm min}}^{p_\chi^{\rm max}} dp_\chi  \dfrac{p_\chi}{ E_\chi}\dfrac{d\phi_\chi}{dE_\chi} \dfrac{d\sigma}{dE_R}\,.
\end{equation}
Here, the lower integration boundaries of $p_\chi$ is given through $  p_\chi^{\rm min} =\sqrt{(E_\chi^{\rm min})^2 - m_\chi^2}$ with
\begin{align}
     E_\chi^{\rm min} &=\dfrac{E_R}{2}+\frac{1}{2 m_e}\sqrt{m_e(E_R +2m_e)(E_R m_e +2 m_\chi^2)} ,
\end{align}
and the upper boundary  as before.
The expected number of events is given by,
\begin{equation}
    N_{\rm sig}^{\rm DR} = \int_{E_{\rm th}}^{E_R^{\rm max}} dE_R \, \dfrac{dR}{dE_R}\,, 
\end{equation}
where $E_{\rm th}$ is the threshold recoil energy.
The maximal recoil energy $E_R^{\rm max}$ is either given by the energy range of the  experimental data or half of the progenitor mass. 

For large enough $m_X$, the DR is energetic enough to generate $\mathcal{O}({\rm keV}\text{--}{\rm MeV})$ NR in direct detection and neutrino experiments.
The framework for NR is the same as scattering on free electrons discussed above, but  the recoil cross section become target-dependent; 
see~\cite{Chu:2018qrm} for detailed formulas of the nuclear recoil cross section.

\section{Results}
\label{sec:result}

\subsection{Constraints on the effective couplings}

\begin{figure*}[t]
\begin{center}
\includegraphics[width=0.49\textwidth]{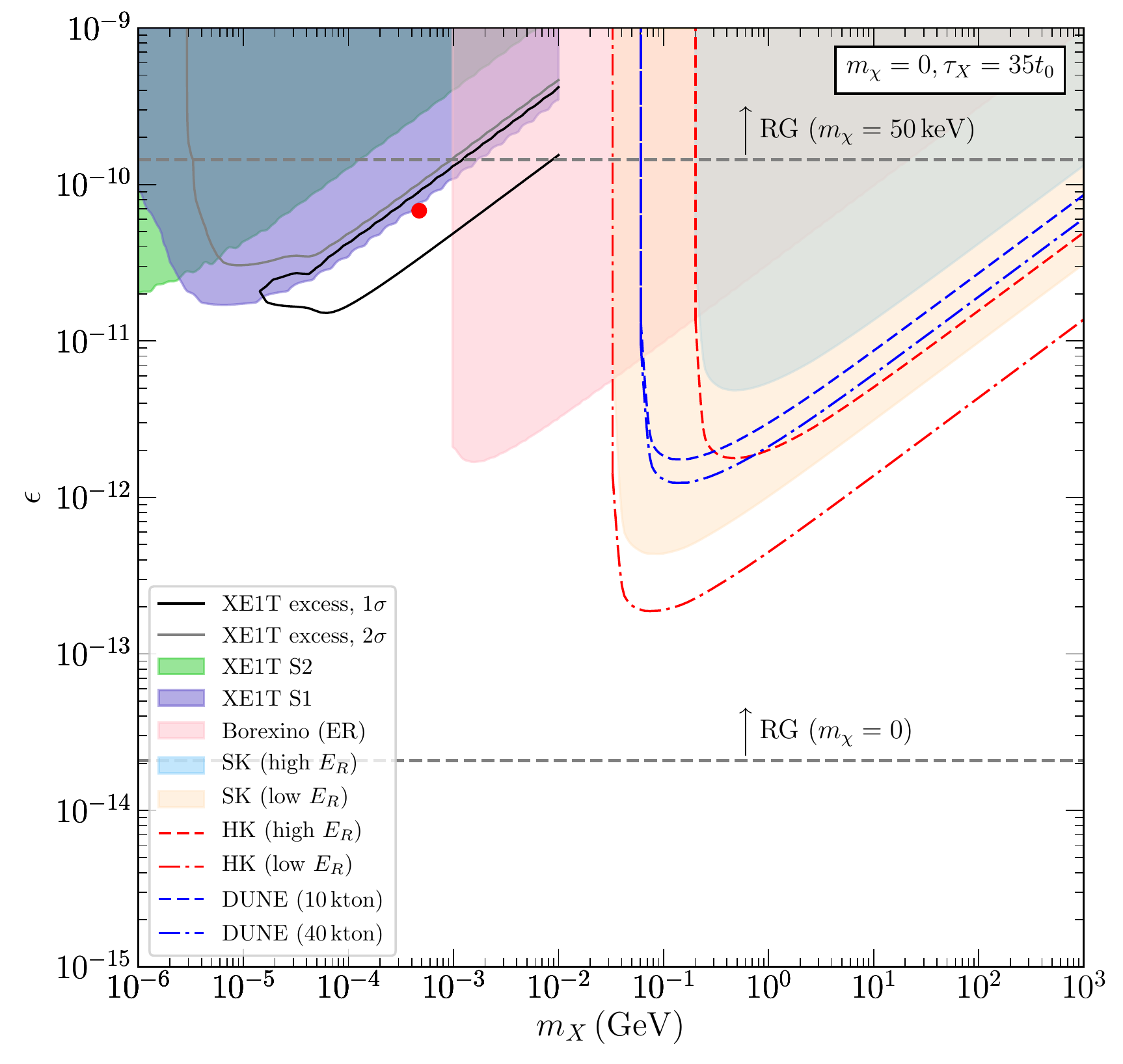}
\includegraphics[width=0.49\textwidth]{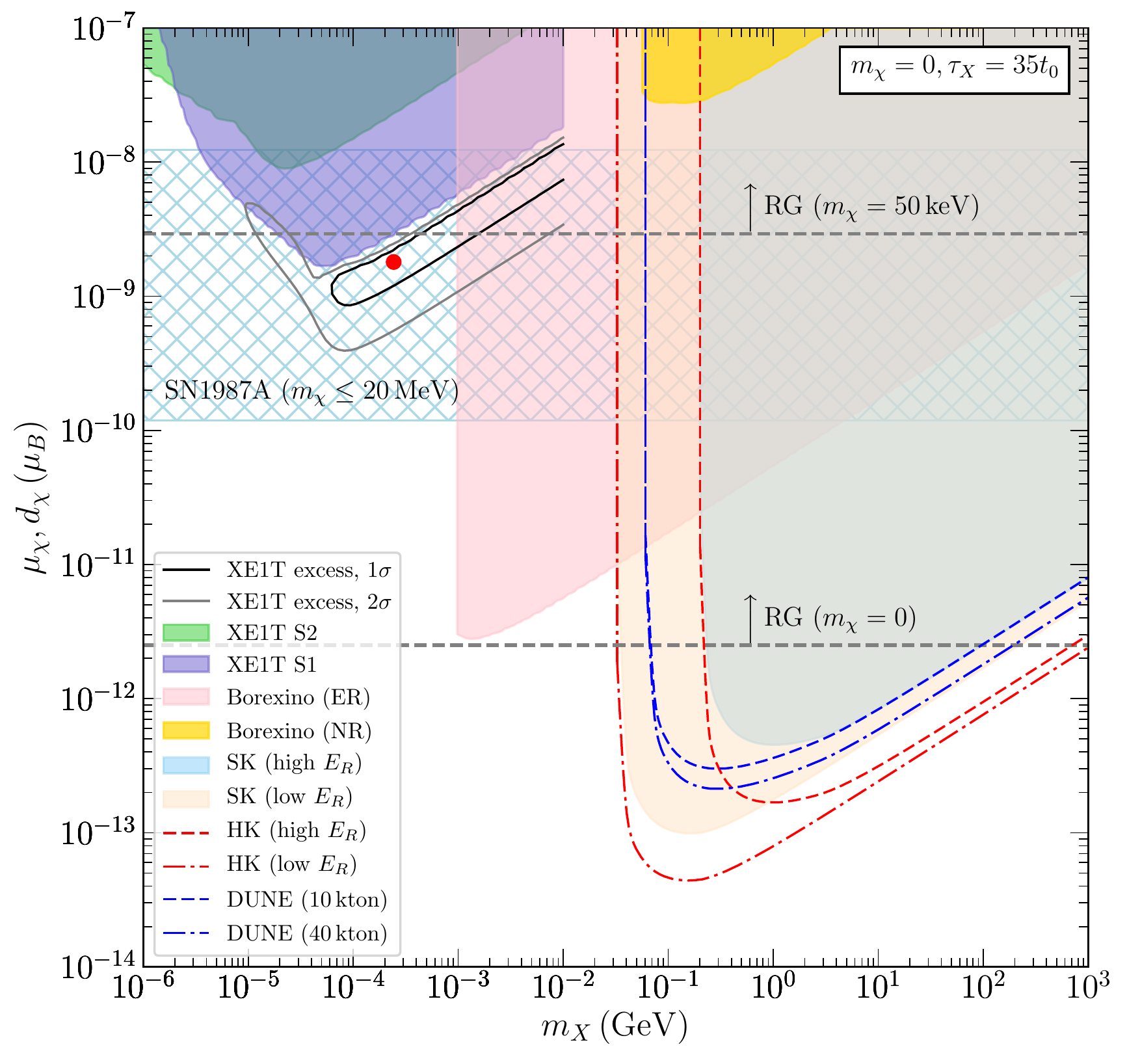}
\end{center}
\caption{Constraints and  forecasts of sensitivity on the  mQ (left panel) and effective MDM/EDM (right panel) interaction of the DR.
In addition, the best-fit values (indicated by the red dots) and the favoured regions explaining the XENON1T excess are shown.
The strongest bounds in the literature, taken from~\cite{Davidson:2000hf,Chu:2019rok}, from the anomalous energy loss inside red giant stars are included for comparison; their strength depends on the DR mass.
For dimension~5 operators, we also show the constraints from the anomalous cooling of SN1987A and taken from~\cite{Chu:2019rok}.
}
\label{Fig:bound}
\end{figure*}

\begin{figure}[tb]
\begin{center}
\includegraphics[width=0.49\textwidth]{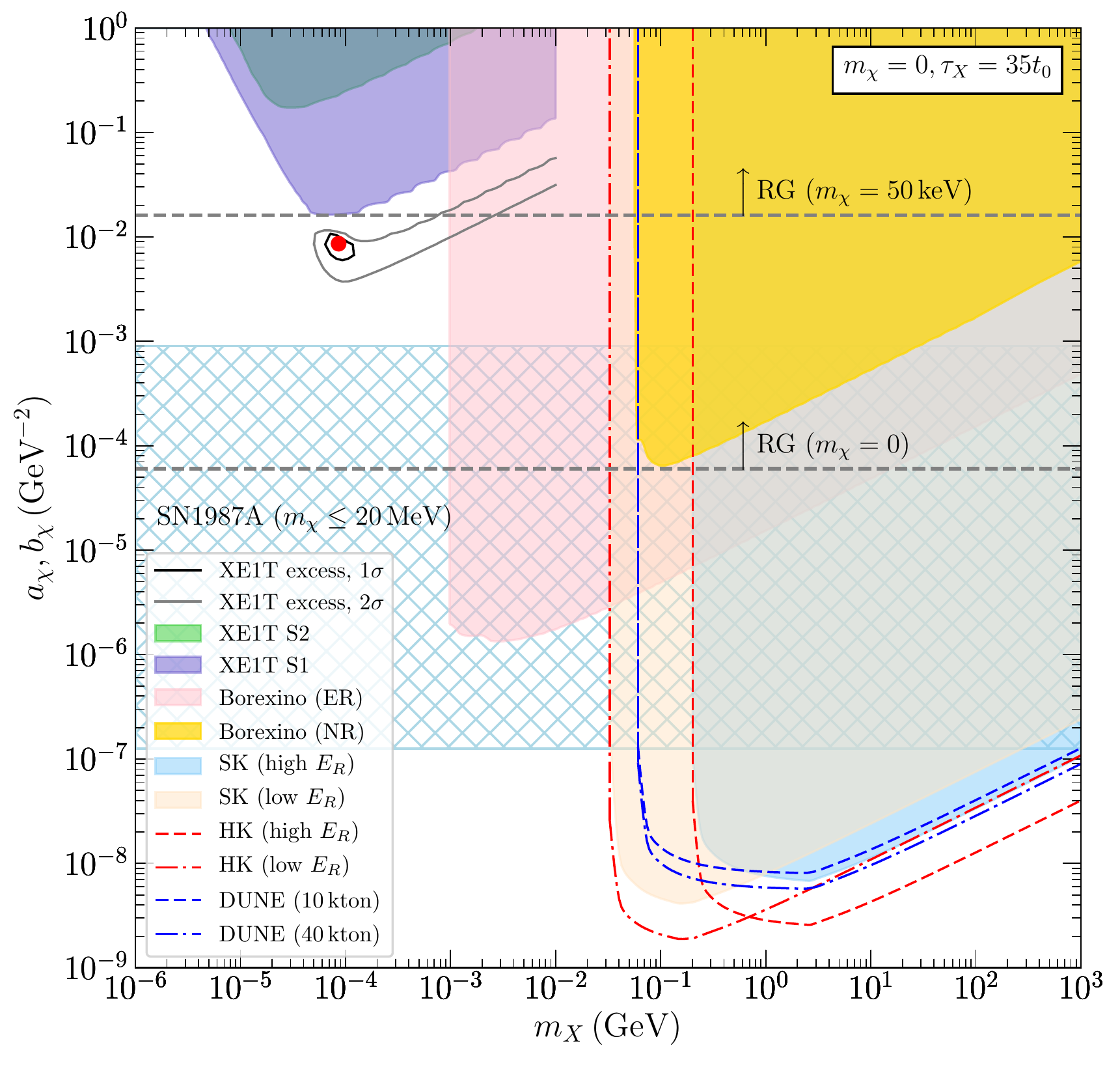}
\end{center}
\caption{The constraints and the forecasts of sensitivity on the effective AM/CR interaction as labeled.
The hatched region shows the anomalous cooling constraint from SN1987A and the horizontal dashed lines are constraints from the energy loss inside RG stars~\cite{Chu:2019rok}.
}
\label{Fig:bound_dim6}
\end{figure}

We show the resulting constraints (shaded regions) and  forecasts of sensitivity (lines) for millicharged DR in the left panel of Fig.~\ref{Fig:bound}, for MDM/EDM interactions in the right panel of Fig.~\ref{Fig:bound} and for AM/CR interactions in Fig.~\ref{Fig:bound_dim6}.
We also show the XENON1T excess favored region, with details on the  fitting procedure  given in Sec.~\ref{Sec:XE1T_excesss}.
Previous constraints derived from the anomalous energy loss in red giant stars (RG) and SN1987A cooling are included for comparison~\cite{Davidson:2000hf,Chu:2019rok}, which apply when $m_\chi$ is smaller than the plasma frequency in the stellar environment: $m_\chi \leq 10\,{\rm keV} (20\,{\rm MeV})$ for RG (SN1987A); see also~\cite{Chang:2018rso,Chang:2019xva}.
For mQ, we note that there exist additional bounds from galaxy cluster magnetic fields~\cite{Kadota:2016tqq} and the timing of radio waves~\cite{Caputo:2019tms}.
However, both of them scale with the DR mass, thus they are not included in the figures. 

Due to the energy dependence in the cross sections, the experiments with higher threshold are more important for higher dimensional operators. 
We see that current SK and future HK and DUNE can all provide better sensitivity than current best limit from the stellar energy loss for dimension~5 and~6 operators, assuming $\tau_X = 35 t_0$.
For mQ, the improvement of sensitivity between experiments that probe free electron scattering and XENON1T is not so notable compared to higher dimensional operators. 

The constraints derived in this paper for $m_\chi = 0$ also apply to electromagnetic form factors of neutrinos if they play the role of DR.
On the other hand, if only taking the SM weak interactions of neutrinos,  direct detection and neutrino experiments put bounds on the mass and lifetime of the progenitor $X$; see, \textit{e.g.}, Ref.~\cite{Cui:2017ytb}.

\subsection{XENON1T excess}
\label{Sec:XE1T_excesss}

\begin{figure}[tb]
\centering
\includegraphics[width=\columnwidth]{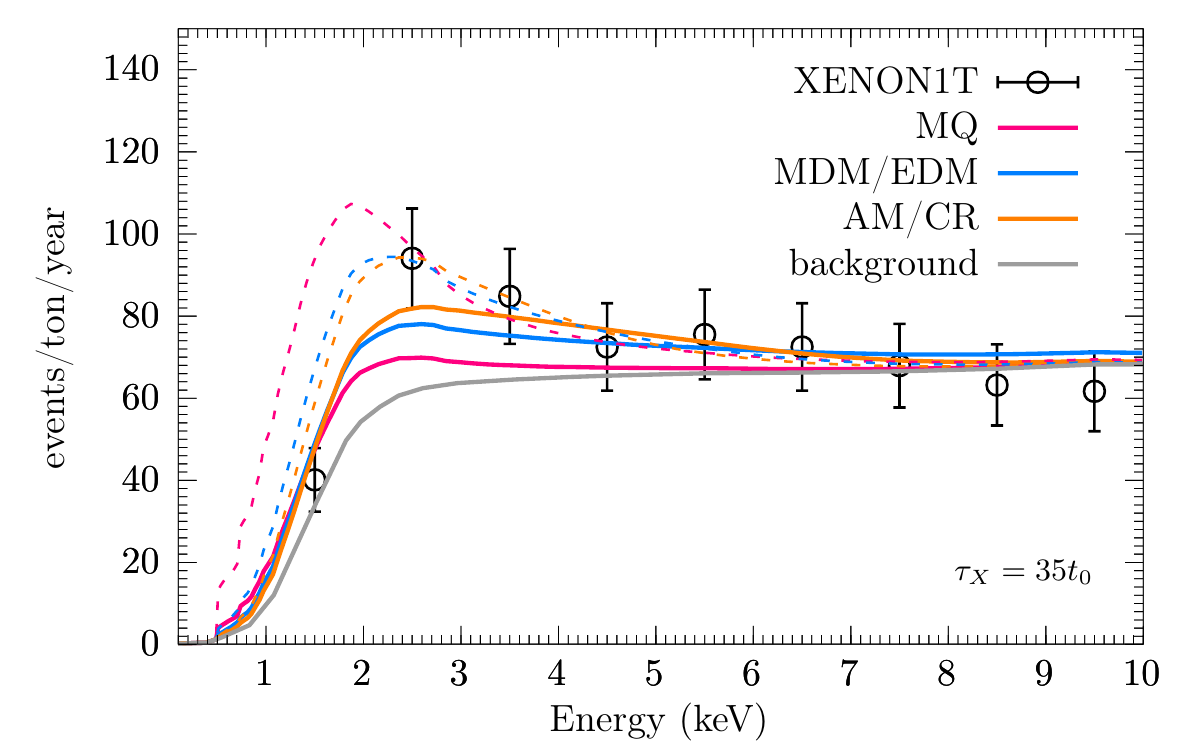}
\caption{Best-fit event rate to the XENON1T excess for each effective electromagnetic interaction. We demonstrate cases including/excluding the first bin in the fit by the solid/dashed lines; the event rates are summed with the background expectation given by the gray solid line.}
  \label{fig:XE1T_fit}
\end{figure}

\begin{table}[t]
 \centering
 \begin{tabular}{lccc}
	 \toprule
   & $m_X\,({\rm keV})$& coupling & $\chi^2/{\rm dof}$ \\
  \midrule
  mQ  & $472$ & $\epsilon = 6.8\times 10^{-11}$  & $9.2/7$ \\
  \quad excl.~first bin & $183$ & $\epsilon = 8.9\times 10^{-11}$  & $1.5/6$ \\
  MDM, EDM  & $243$ & $\mu_\chi,d_\chi = 1.8\times 10^{-9}\,\mu_B$  & $5.8/7$ \\
   \quad excl.~first bin & $81$ & $\mu_\chi,d_\chi = 1.8\times 10^{-9}\,\mu_B$  & $1.1/6$ \\
    AM, CR  & $86$ & $a_\chi,b_\chi = 8.6\times 10^{-3}\,{\rm GeV}^{-2}$  & $3.6/7$ \\
   \quad excl.~first bin & $71$ & $a_\chi,b_\chi = 1.1\times 10^{-2}\,{\rm GeV}^{-2}$  & $1.1/6$ \\
 \bottomrule
 \end{tabular}
 \caption{Best-fit values of $m_X$ and  strength of the electromagnetic interaction as well as the corresponding $\chi^2/{\rm dof}$. A lifetime of $\tau_X=35t_0$ is assumed.}
  \label{tab:bestfit}
 \end{table}

 \begin{figure*}[t]
\begin{center}
\includegraphics[width=0.32\textwidth]{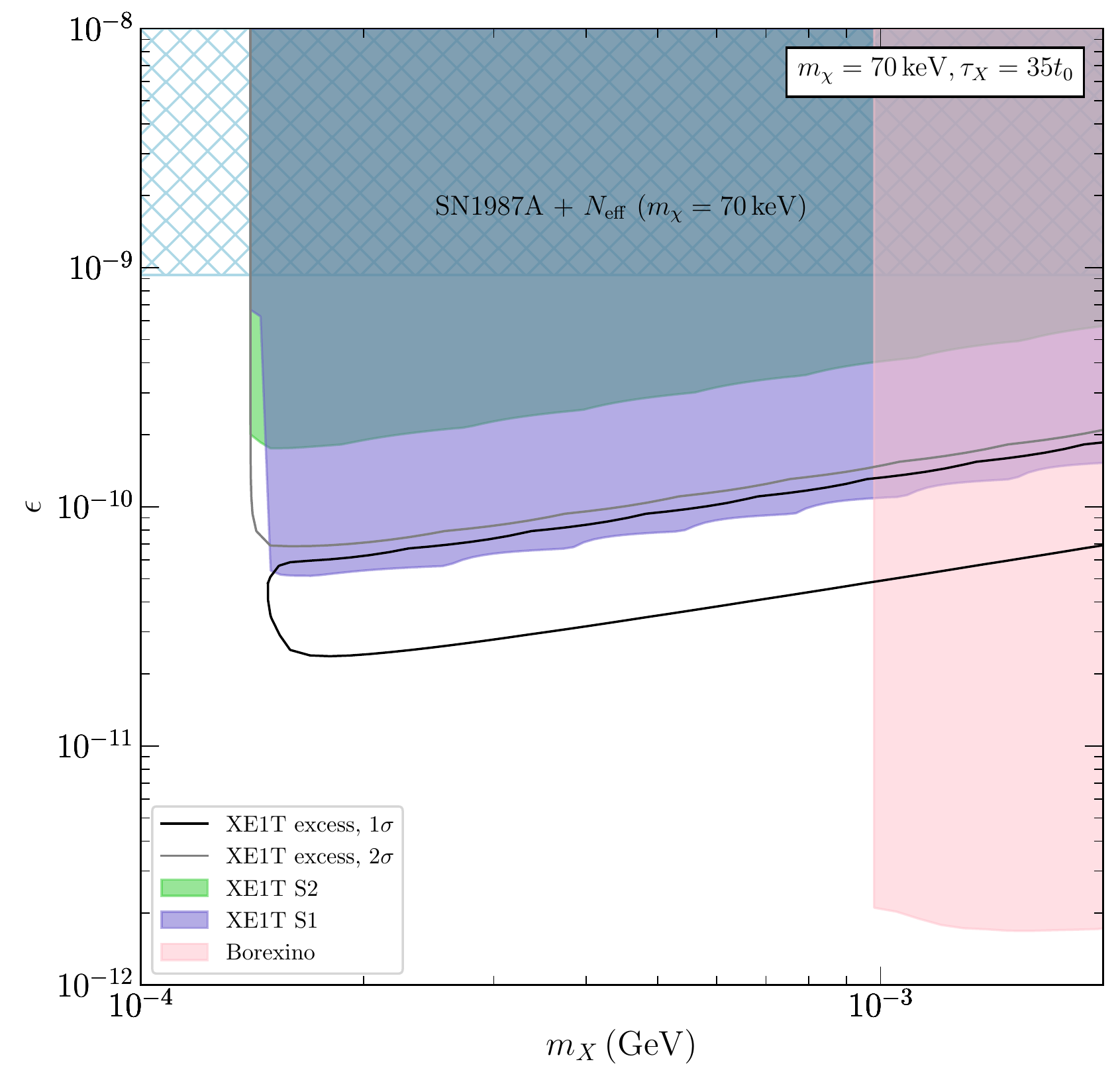}
\includegraphics[width=0.32\textwidth]{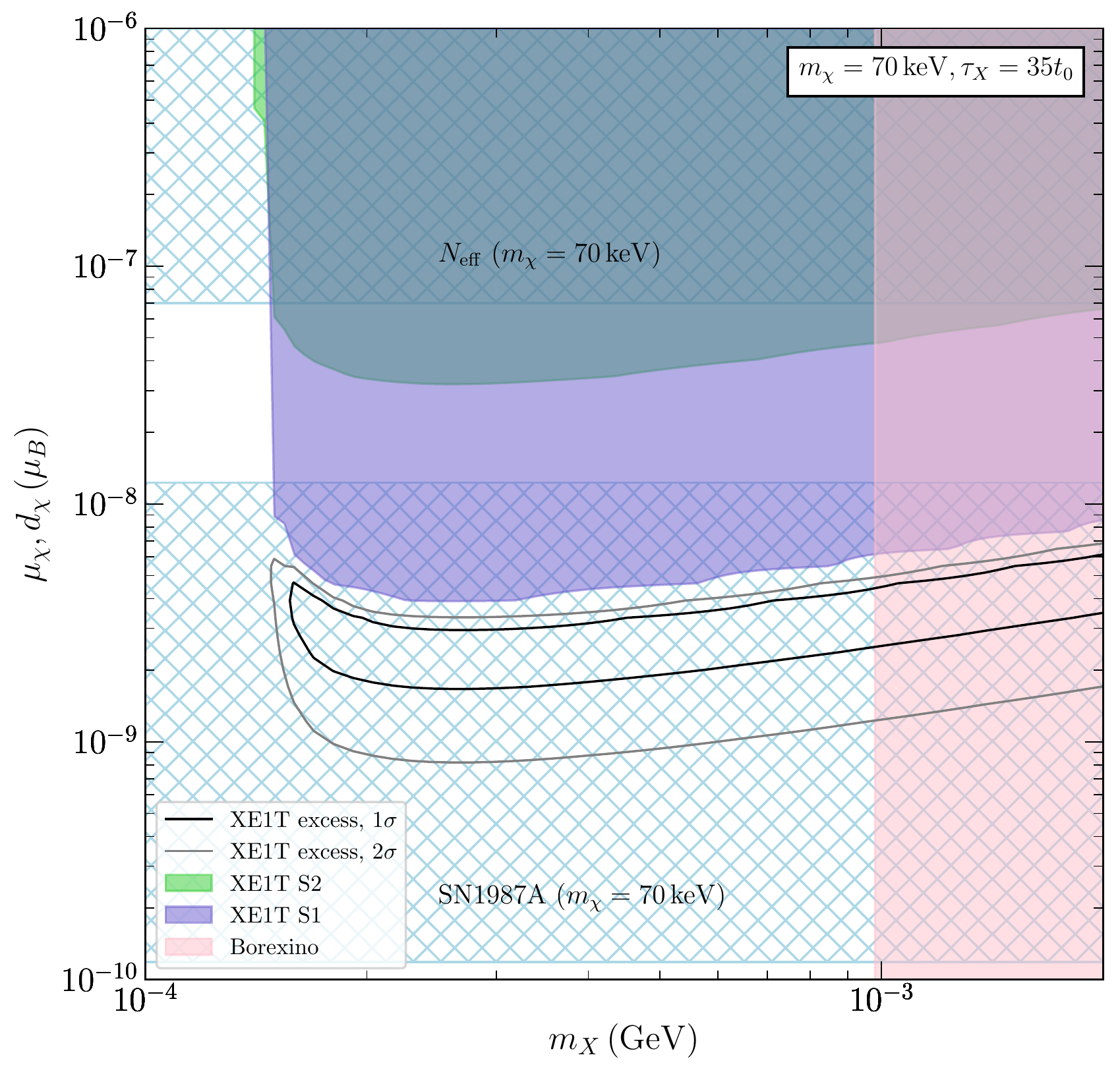}
\includegraphics[width=0.32\textwidth]{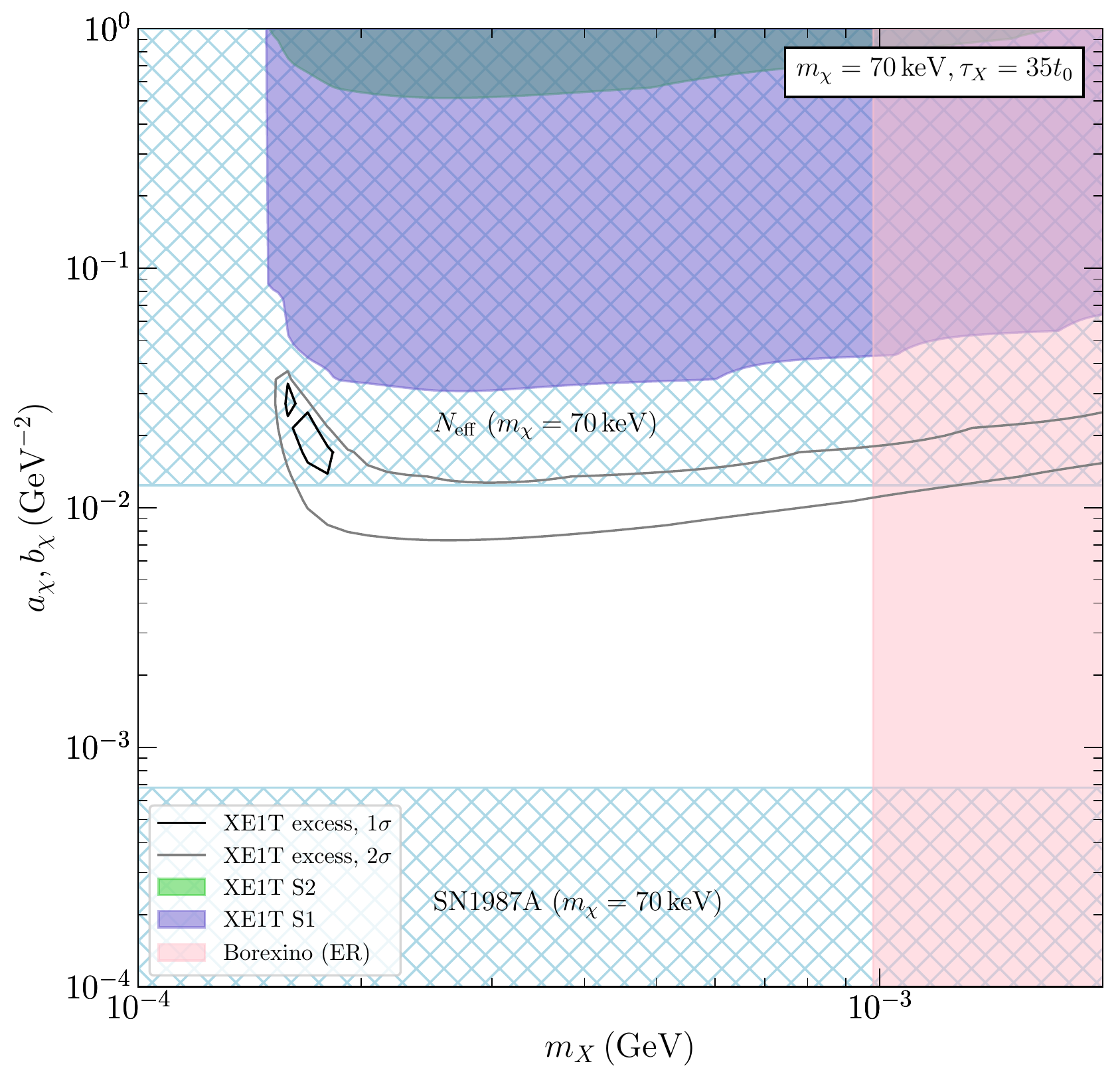}
\end{center}
\caption{Zoom-in figures with massive DR ($m_\chi = 70\,{\rm keV}$) for the parameter space favoured by the XENON1T excess. {\em Left panel (mQ)}: the parameter space for explaining the XENON1T excess is not constrained by either stellar energy loss arguments or $N_{\rm eff}$. {\em Middle panel (MDM/EDM)}: the parameter space is fully covered by the SN1987A bound. {\em Right panel (AM/CR)}: part of the parameter space is ruled out by $N_{\rm eff}$. The SN1987A and $N_{\rm eff}$ constraints are adopted from~\cite{Davidson:2000hf,Chu:2019rok}. Here we neglect the differences between EDM and MDM as well as between AM and CR operators as they are not resolved except at the very kinematic endpoint $m_X\simeq 140\,\keV$ as DR remains (semi-)relativistic everywhere else. 
}
\label{Fig:70keV}
\end{figure*}

In light of the recent excess in the $\mathcal{O}({\rm keV})$ recoil energy range observed by XENON1T~\cite{Aprile:2020tmw}, we also explore the possibility of explaining the excess with DR. assuming the background modelling is correct.
This lines up with  several other new physics scenarios and their constraints that have been investigated in this context~\cite{Aprile:2020tmw,Boehm:2020ltd,Takahashi:2020bpq,An:2020bxd,Khan:2020vaf,Bloch:2020uzh,DiLuzio:2020jjp,*Buch:2020mrg,*Gao:2020wer,*Dent:2020jhf,*Dessert:2020vxy,*Sun:2020iim,*Cacciapaglia:2020kbf,*Croon:2020ehi,*Li:2020naa,*Athron:2020maw,*Millea:2020xxp,*Arias-Aragon:2020qtn,*Alonso-Alvarez:2020cdv,*Choi:2020kch,*Chiang:2020hgb,*Kannike:2020agf,*Fornal:2020npv,*Su:2020zny,*Chen:2020gcl,*Cao:2020bwd,*Jho:2020sku,*DelleRose:2020pbh,*Alhazmi:2020fju,*Basu:2020gsy,*Davoudiasl:2020ypv,*Choudhury:2020xui,*Ema:2020fit,*Cao:2020oxq,*Bally:2020yid,*AristizabalSierra:2020edu,*Ge:2020jfn,*Coloma:2020voz,*Miranda:2020kwy,*Babu:2020ivd,*Shoemaker:2020kji,*Arcadi:2020zni,*Karmakar:2020rbi,*Harigaya:2020ckz,*Bell:2020bes,*Lee:2020wmh,*Bramante:2020zos,*An:2020tcg,*Chao:2020yro,*Baek:2020owl,*He:2020wjs,*Borah:2020jzi,*Du:2020ybt,*Choi:2020udy,*Paz:2020pbc,*Dey:2020sai,*Budnik:2020nwz,*Zu:2020idx,*Lindner:2020kko,*McKeen:2020vpf,*Gao:2020wfr,*Ko:2020gdg,*Okada:2020evk}.%
\footnote{Reference~\cite{Khan:2020vaf} considers all form factors of~\eqref{eq:Lagrangians} but for SM neutrinos, fitting the excess with their solar flux. The required interaction strengths are excluded from stellar energy loss constraints~\cite{Chu:2019rok}, disfavoring the ``solar option'' altogether.}%
Moreover, PandaX-II reports for its own data  that it is both, consistent with a new physics contribution as well as with a fluctuation of background~\cite{Zhou:2020bvf}.
Thus the observational status of an excess in XENON1T remains unclear at the moment. 

In Fig.~\ref{fig:XE1T_fit}, we show the best-fit event rate induced by DR and the data in the energy range $[0,10]\,{\rm keV}$ in two fitting scenarios, including and excluding the first bin.
By excluding the first bin, the recoil spectrum can better fit to the peak of excess, but at the expense of significantly overshooting the first bin, (See also related discussions in Refs. \cite{Bloch:2020uzh,Harnik:2020ugb}).
When the first bin is included in the fit, the second bin cannot be filled but the overall fit is still satisfactory, similar to the anomalous neutrino magnetic dipole moment explanation~\cite{Aprile:2020tmw,Boehm:2020ltd}.
The corresponding best-fit parameters and $\chi^2/{\rm dof}$ are given in Table.~\ref{tab:bestfit}.
We observe that higher-dimensional operators yield improved fits to the excess, as their recoil spectra are less peaked at low $E_R$. 
We also note that the best-fit coupling of dimension~5 operator is consistent with the best-fit anomalous magnetic dipole moment of neutrino~\cite{Boehm:2020ltd},~\textit{i.e.}, 
\begin{equation}
    \phi_\chi^{\rm best} \times \left(\mu_\chi^{\rm best} \right)^2 \simeq \phi_\nu^{\rm solar} \times \left(\mu_\nu^{\rm best} \right)^2\,,
\end{equation}
although the free electron approximation is adopted in~\cite{Boehm:2020ltd}.

For massless DR, the favoured parameter space for the excess is excluded by stellar energy loss constraints, such as red giant stars for dimension~4 and~5 operators and SN1987A for the dimension~6 operators, shown in Fig.~\ref{Fig:bound}, and taken from~\cite{Davidson:2000hf,Chu:2019rok}; see also~\cite{Chang:2018rso,Chang:2019xva}.
However, stellar energy loss is effective only when $\chi$ production is kinematically allowed.  Taking DR with $m_\chi = 70\,{\rm keV}$, the constraints from the stellar energy loss are alleviated, while, at the same time, leaving the fits to the XENON1T excess to remain unchanged. This is owed to the relatvistic nature of considered DR. Finally, we consider the constraint from the measured number of relativistic degrees of freedom $N_{\rm eff}$, as $\chi$ particles are also populated in the early universe through plasmon decay and electron-positron annihilation~\cite{Davidson:2000hf,Chu:2019rok,Chang:2019xva}.
As shown in Fig.~\ref{Fig:70keV} for mQ (left panel), MDM/EDM (middle panel) and AM/CR (right panel),
 there remains allowed parameter space  for explaining the XENON1T excess for dimension~4 and~6 operators.
For dimension~5 operators, the viable parameter space is covered by the SN1987A bound. 

\section{Conclusion}
\label{sec:conclusions}

In this paper we have considered the possibility that DM $X$ is unstable and decays to a pair $\bar \chi \chi$ which itself couples to the SM through effective interactions mediated by the photon. We consider the possibility of millicharge of $\chi$, magnetic and electric dipole moments, and the less familiar anapole moment and charge radius interaction. 
The emerging DR flux from DDM is then probed in underground rare-event searches. For $m_X \lesssim 1~\MeV$ direct detection experiments offer the best sensitivity with their ability of registering keV-scale  energy depositions and below. Heavier progenitors are better probed with neutrino experiments, as $\chi$-induced events leave MeV-scale signals. For concreteness, in this work we have chosen a benchmark value of $\tau_X = 35 t_0$ with the bulk of DM still to decay in the distant future.

The scattering of $\chi$ on electrons is the most important signal channel. 
We demonstrate that the recent (S1+S2) data from the XENON1T experiment yields $\epsilon\lesssim 2\times 10^{-11}$ at $m_X\simeq 10~\keV$, and  
$d_\chi,\mu_\chi \lesssim 2\times 10^{-9}\mu_B$ as well as  $a_\chi,b_\chi \lesssim 2\times 10^{-2}\GeV^{-2}$ at $m_X\simeq 100\,\keV$. In addition, we find that it is also possible to reach a satisfactory fit to the reported excess of events seen in the XENON1T data at few~keV energy. The fit improves by increasing the dimensionality of the operator, as the lowest energy bin in the data prohibits too strong of an  IR-biased signal. The AM/CR interaction thereby yields the best fit. 
The DR mass-dependence is relatively mild in those drawn conclusions  as these particles retain their (semi-)relativistic nature except at the very kinematic edge $2m_\chi \simeq m_X$. However, stellar and cosmological constraints critically depend on $m_\chi$. By choosing a benchmark value of $m_\chi = 70\,\keV$ we demonstrate that a XENON1T explanation remains intact for mQ and for the dim-6 AM and CR operators, evading bounds from the anomalous energy loss inside RG stars, of the proto-neutron star of SN1987A and from the cosmological  $N_{\rm eff}$ limit.

For progenitor masses $m_X\gtrsim 1\,\MeV$ Borexino has the best sensitivity reaching $\epsilon\lesssim 10^{-12}$,  
$d_\chi,\mu_\chi \lesssim 3\times 10^{-12}\,\mu_B$ and  $a_\chi,b_\chi \lesssim 2\times 10^{-6}\,\GeV^{-2}$ at $m_X\simeq 1\,\MeV$. These limits rely on the detailed modeling of Borexino backgrounds and its solar neutrino-induced events. The limits are eventually surpassed by the ones from SK, once DR induces electron recoils above the solar neutrino endpoint energies. Best sensitivity is attained for $m_X \simeq 100\,\MeV$ with  $\epsilon\lesssim 4\times 10^{-13}$,  
$d_\chi,\mu_\chi \lesssim  10^{-13}\,\mu_B$ and  $a_\chi,b_\chi \lesssim 4\times 10^{-9}\,\GeV^{-2}$. Finally, we also provide forecasts for HK and DUNE, with relatively mild expected improvements.

\vspace{.3cm}

\paragraph*{Acknowledgments}
JLK is supported by the Austrian Science Fund FWF under the Doctoral Program W1252-N27 Particles and Interactions. JP is supported by the New Frontiers program of the Austrian Academy of Sciences.  MP is supported in part by U.S. Department of Energy (Grant No. desc0011842).
 We acknowledge the use of computer packages for
  algebraic calculations~\cite{Mertig:1990an,Shtabovenko:2016sxi}. 

\appendix

\section{Recoil cross section on bound electron}
\label{App:xsec_electron}
For scattering on bound electrons, the differential cross section for massive DR that may either be relativistic or non-relativistic reads~\cite{An:2017ojc},
\begin{equation}
\begin{split}
\dfrac{d\sigma v }{ dq dE_R }\bigg|_{(n,l)} &= \dfrac{\bar{\sigma}_e}{8\mu_{\chi e}^2 E_R }\dfrac{m_\chi^2 }	{p_\chi E_\chi}\int d\Omega_{\vec{p}'_e} \,q |f_{n,l} (\vec{q})|^2  \\
&\quad\,\,\times |F_\chi (q, E_\chi)|^2\,,
\end{split}
\end{equation}
where $E_R$ is the electron recoil energy, $v$ is the relative velocity, $\mu_{\chi e}$ is the reduced mass between $\chi$ and $e^-$, $q$ is the momentum transfer, $d\Omega_{\vec{p}'_e}$ is the solid angle element of the final state electron and $|f_{n,l} (\vec{q})|^2 $ is the atomic form factor for the atomic state $(n,l)$~\cite{Essig:2011nj,Essig:2012yx}\footnote{Note that often  $|f_{\rm ion} (q)|^2 = \int d\Omega_{\vec{p}'_e}\,|f_{n,l}(\vec{q})|^2$ is written in the literature.}.
In this work, we use the numerical result of $|f_{n,l} (\vec{q})|^2 $ that was obtained from an atomic calculation described in~\cite{Essig:2019xkx}.
We normalize our results to the non-relativistic effective scattering cross section $\bar{\sigma}_e$ on a free electron, evaluated at a typical atomic momentum transfer $q_0 \simeq \alpha m_e$ and at vanishing kinetic energy, $E_\chi = m_\chi$,
\begin{equation}
\label{eq:barsigmae}
	\bar{\sigma}_e = \dfrac{\mu_{\chi e}^2}{16\pi m_\chi^2 m_e^2} \overline{|\mathcal{M}(q= q_0, E_\chi = m_\chi )|}^2\,,
\end{equation}
where $\overline{|\mathcal{M}|}^2$ is the squared amplitude,  averaged over the initial state spins and summed over the final state spins.
We assume $m_e$ is much larger than the momentum transfer $q$ and the deposited energy $\Delta E$, which is  valid when we consider the scattering with bound electrons in this work. We then expand $\overline{|\mathcal{M}(q,E_\chi)|}^2$ in $q/m_e = O(\alpha)$ and are careful to additionally retain the leading terms in a velocity-expansion that become relevant in the non-relativistic limit $v < \alpha$. For mQ, EDM, CR the leading order terms in both expansions  coincide; for MDM and AM, the leading order term in $q/m_e$ is velocity suppressed in the non-relativistic limit, and we add the term that is not velocity suppressed but of higher order in $q/m_e$. We find
\begin{align}
\label{Eq:DR_xsec_mQ}
{\rm mQ}: \bar{\sigma}_e &= \epsilon^2 \pi \alpha^2 \dfrac{16 m_\chi^2 m_e^2 }{q_0^4 (m_\chi + m_e)^2}\,, \\
{\rm MDM}: \bar{\sigma}_e &= \mu_\chi^2 \alpha \dfrac{  m_\chi^2 }{(m_\chi + m_e)^2}\,, \\
{\rm EDM}: \bar{\sigma}_e &= d_\chi^2 \alpha \dfrac{ 4m_\chi^2 m_e^2}{q_0^2 (m_\chi + m_e)^2}\,, \\
{\rm AM}: \bar{\sigma}_e &= a_\chi^2 \alpha \dfrac{q_0^2 m_\chi^2 }{(m_\chi + m_e)^2} \,, \\
{\rm CR}: \bar{\sigma}_e &= b_\chi^2 \alpha \dfrac{ 4m_\chi^2 m_e^2  }{(m_\chi + m_e)^2}\,,
\end{align}
where $\epsilon$ is the millicharge of $\chi$ in units of the elementary charge $e$, $\mu_\chi$ and $d_\chi$ are the magnetic and electric dipole moment of $\chi$, and $a_\chi$ and $b_\chi$ are the anapole moment and charge radius coupling of $\chi$.
The listed non-relativistic effective scattering cross sections agrees with the ones found in~\cite{Essig:2011nj,Chu:2018qrm}.

The dark matter form factor is defined as 
\begin{align}
	|F_\chi (q, E_\chi)|^2 = \dfrac{\overline{|\mathcal{M}(q, E_\chi)|}^2}{\overline{|\mathcal{M}(q= q_0, E_\chi = m_\chi )|}^2} \, ,
\end{align}
with the concrete expressions for the respective effective operators given by
\begin{align}
\label{Eq:DR_form_mQ}
    {\rm mQ}: |F_\chi (q, E_\chi)|^2 &= \dfrac{E_\chi^2 q_0^4 }{m_\chi^2 q^4}\,, \\
    \label{Eq:DR_form_MDM}
{\rm MDM}:|F_\chi  (q, E_\chi)|^2 &=\dfrac{4m_e^2 \left( E_\chi^2 -m_\chi^2 \right)}{q^2 m_\chi^2 }+ 1\,, \\
\label{Eq:DR_form_EDM}
{\rm EDM}:|F_\chi  (q, E_\chi)|^2 &= \dfrac{E_\chi^2 q_0^2 }{m_\chi^2 q^2}\,, \\
\label{Eq:DR_form_AM}
{\rm AM}:|F_\chi  (q, E_\chi)|^2 &= \dfrac{4 m_e^2 \left(E_\chi^2 -m_\chi^2 \right) + q^2 m_\chi^2 }{q_0^2 m_\chi^2}\,, \\
\label{Eq:DR_form_CR}
{\rm CR}:|F_\chi  (q, E_\chi)|^2 &= \dfrac{E_\chi^2}{m_\chi^2}\,.
\end{align}
Unlike in the direct detection literature that is concerned with chiefly non-relativistic scatterings, the form factors defined here carry an additional dependence on $E_\chi$ as is generally the case for relativistic scattering processes. 
When taking the non-relativistic limit $E_\chi \simeq m_\chi$, we retrieve the non-relativistic dark form factors found in the literature~\cite{Essig:2011nj,Chu:2018qrm}.
In addition, in the relativistic limit, $E_\chi \gg m_\chi$, the helicity suppression introduced by $\gamma^5$ drops out, and the respective dim-5 and dim-6 form factors become equivalent. 
The dark form factors presented here are applicable across the entire kinematic regime.

For massless DR, $m_\chi = 0$, the differential cross section can be written as,
\begin{equation}
    \dfrac{d\sigma v }{ dq dE_R }\bigg|_{(n,l)} = \dfrac{\bar{\sigma}_e}{8 E_R p_\chi^2} \int d\Omega_{\vec{p}'_e}\,q |f_{n,l} (\vec{q})|^2 |F_\chi (q, p_\chi)|^2\,.
\end{equation}
It should be noted that $\bar \sigma_e$ in~\eqref{eq:barsigmae} is ill-defined for $m_\chi\to 0$, but in the product $\bar{\sigma}_e|F_\chi  (q, E_\chi)|^2$ the mass-dependence cancels out. Hence, in practice, keeping with the usually adopted convention for the definition of $\bar \sigma_e$ does not pose any obstruction.

\section{Recoil cross section on free particle}
\label{App:xsec_free}

In agreement with the previous work on the dark sector-photon interactions~\cite{Chu:2018qrm}, we list here for completeness the recoil cross sections for scattering on free electrons, 
\begin{widetext}
\begin{align}
{\rm mQ}: \dfrac{d\sigma}{dE_R} &=\epsilon^2 \pi \alpha^2 \dfrac{m_e (E_R^2 +2 E_\chi^2)-E_R (2 E_\chi m_e +m_e^2 +m_\chi^2)}{(E_\chi^2 -m_\chi^2)E_R^2 m_e^2}\,, \\
{\rm MDM}: \dfrac{d\sigma}{dE_R} &= \mu_\chi^2 \alpha \dfrac{(E_R - 2m_e) m_\chi^2 - 2(E_R -E_\chi) E_\chi m_e}{2(E_\chi^2 -m_\chi^2 )E_R m_e}\,, \\
{\rm EDM}: \dfrac{d\sigma}{dE_R} &= d_\chi^2 \alpha \dfrac{  2E_\chi m_e (E_\chi - E_R) - E_R m_\chi^2 }{2 (E_\chi^2 - m_\chi^2) E_R m_e}\,, \\ 
{\rm AM}: \dfrac{d\sigma}{dE_R} &= a_\chi^2 \alpha \dfrac{m_e \left[ E_R^2 -E_R (2E_\chi +m_e) +2 E_\chi^2 \right]+m_\chi^2 (E_R -2m_e)}{E_\chi^2 -m_\chi^2} \,, \\
{\rm CR}: \dfrac{d\sigma}{dE_R} &= b_\chi^2 \alpha \dfrac{E_R^2 m_e -E_R (2E_\chi m_e +m_e^2 +m_\chi^2) + 2E_\chi^2 m_e}{E_\chi^2 -m_\chi^2}\,.
\end{align}
\end{widetext}
For the general expressions that are applicable for the analogous scattering on nuclei, see App.~E in~\cite{Chu:2018qrm}.

\bibliography{refs}

\end{document}